\documentclass{appolb}
\usepackage{epsfig}
\newcommand{\Lqcd}{\Lambda_{\rm QCD}}
\newcommand{\bea}{\begin{eqnarray}}
\newcommand{\nn}{\nonumber}
\newcommand{\eea}{\end{eqnarray}}
\newcommand{\be}{\begin{equation}}
\newcommand{\ee}{\end{equation}}
\newcommand{\bi}{\begin{itemize}}
\newcommand{\ei}{\end{itemize}}
\newcommand{\Br}{{\rm Br}}
\newcommand{\RE}{{\rm Re}}
\newcommand{\IM}{{\rm Im}}

\begin{document}


\title{Search for New Physics in Electroweak Penguins via $B_s$ Decays
\thanks{Presented at FLAVIAnet Topical Workshop ``Low energy constraints on extensions
of the Standard Model'', Kazimierz, Poland, 23-27 July 2009.}%
}

\author{L. Hofer, D. Scherer
\address{
Institut f\"ur Theoretische Teilchenphysik, \\
Karlsruhe Institute of Technology, Universit\"at Karlsruhe, \\
D--76128 Karlsruhe, Germany}
\and
L. Vernazza
\address{Institut f\"ur Theoretische Physik E,
RWTH Aachen University,\\
D--52056 Aachen, Germany}
}

\maketitle

\begin{abstract}
The discrepancies found in the $\bar{B}\to \pi \bar{K}$
decays between theory and experiment suggest the presence
of new physics in the electroweak penguin sector of the
theory. We show that this hypothesis can be tested more
efficiently including in the analysis the non-leptonic
decays $\bar{B}_s\to \phi\pi,\phi\rho$.
\end{abstract}

\PACS{13.25.Hw, 12.60.Cn \\[-1.3cm]}
\begin{flushright}
{\small
Preprint: TTP~09-35 \\
PITHA~09/25\\
SFB/CPP-09-88
}
\end{flushright}

\section{Introduction}

In recent years tensions at the $\sim 2$ $\sigma$ level have been found
between theoretical predictions and experimental results in the
$\bar{B}\to\pi \bar{K}$ decays. These differences point in the direction
of new physics (NP) in the electroweak (EW) penguin sector of the theory \cite{NewEW}.
The issue is still open because the discrepancies decreased since
the first time they were found and it is not yet possible to claim for NP.
On the one hand, this is due to the still insufficient experimental statistics,
which however will improve at LHCb and the future super-$B$ factories;
on the other hand, non-leptonic $B$ decays are still a challenge to theory.
Due to the dominant low-energy QCD effects, it is difficult to single
out the high-energy weak transition which is responsible for the decays and
which could contain NP effects. Methods developed so far rely on
flavour symmetries of QCD or on the factorization properties of
low-energy QCD dynamics (QCDF) \cite{QCDF,QCDF2}. Alas, none of the two is
able to predict the decay amplitudes with the required precision.
The former is applicable only to a handful of decays while
the latter, which implies an expansion of the amplitudes in
$\Lqcd/m_b$, receives important contributions from a number of
subleading terms which can only be estimated.
Given this situation, in order to decide whether the tensions in the
$\bar{B}\to\pi\bar{K}$ data are indeed due to NP, it is important to
inspect also other non-leptonic decays, where effects of a new EW
penguin amplitude are expected to be large.
Among these there are the $\bar{B}\to \rho\bar{K},\pi\bar{K}^*,\rho\bar{K}^*$
and $\bar{B}_s\to \phi\pi,\phi\rho$ decays. We focus in particular on
the latter two because they are pure isospin-violating decays whose branching
ratio in the Standard Model (SM) is predicted to be small. We consider
various extensions of the SM where new-physics effects arise
in the EW penguin sector and show that the branching ratios
of the $\bar{B}_s\to \phi\pi,\phi\rho$ decays can easily be
enhanced by a factor 2 to 4 and in this way be in reach of LHCb and
the planned super-$B$ factories.

\section{Analysis of the $\bar{B}\to \pi \bar{K}$ modes}

We start reviewing the current status of the $\bar{B}\to \pi \bar{K}$
decays. We compare the most recent experimental results with the
theoretical predictions obtained within QCDF \cite{QCDF2}. Usually, 
one considers ratios of different branching fractions which exploit 
the isospin symmetries of the decays and present smaller uncertainties. 
In particular, the ratios
\bea\label{1}\nn
R_c&\equiv& 2\frac{\Br(B^-\to\pi^0K^-)
+\Br(B^+\to\pi^0K^+)}{\Br(B^-\to\pi^-K^0)+\Br(B^+\to\pi^+K^0)} \\[0.1cm] \nn
&=&  1.23^{+0.24}_{-0.20}|_{\rm THEO}, \qquad 1.12^{+0.07}_{-0.07}|_{\rm EXP}, \\[0.2cm] \nn
R_n&\equiv& \frac{1}{2}\frac{\Br(\bar{B}^0\to\pi^+K^-)
+\Br(B^0\to\pi^-K^+)}{\Br(\bar{B}^0\to\pi^0\bar{K}^0)+\Br(B^0\to\pi^0K^0)} \\[0.1cm] \nn
&=&  1.22^{+0.28}_{-0.22}|_{\rm THEO}, \qquad 0.99^{+0.07}_{-0.07}|_{\rm EXP}, \\[0.2cm] \nn
R&\equiv&2\frac{\Gamma(\bar{B}^0\to\pi^0\bar{K}^0)
+\Gamma(B^-\to\pi^0K^-)}{\Gamma(B^-\to\pi^-\bar{K}^0)
+\Gamma(\bar{B}^0\to\pi^+K^-)} \\[0.1cm]
&=& 1.03^{+0.03}_{-0.02}|_{\rm THEO}, \qquad 1.07^{+0.05}_{-0.05}|_{\rm EXP},
\eea
have been widely considered in the literature. Nowadays the
discrepancy between theory and experiment has decreased and
(\ref{1}) shows that the results are compatible within the
errors. More interesting are the direct CP-asymmetry difference
\bea\label{2}\nn
\Delta A_{\rm CP} &=& A_{\rm CP}(B^-\to \pi^0 K^-)
-A_{\rm CP}(\bar{B}^0\to \pi^+ K^-) \\
&=&0.026^{+0.053}_{-0.049}|_{\rm THEO},\quad 0.148^{+0.027}_{-0.028}|_{\rm EXP}
\eea
and the time-dependent CP asymmetry
\be\label{3}
S_{CP}(\bar{B}^0\to \pi^0 \bar{K}^0)
=0.80^{+0.06}_{-0.08}|_{\rm THEO},\quad 0.57^{+0.17}_{-0.17}|_{\rm EXP},
\ee
which show a $\sim1.5\,\sigma$ and $\sim1\,\sigma$ discrepancy between the
theoretical and experimental result, respectively. The large experimental
value for $\Delta A_{\rm CP}$ is difficult to explain within the SM using
QCDF, which in general predicts  direct CP asymmetries to be small. The
observed data can be accommodated better introducing a new EW amplitude
in such a way that the $\bar{B}\to \pi \bar{K}$ amplitudes read
\bea\label{4}\nn
{\cal A}_{B^-\to \pi^- \bar{K}^0}&=& P\,
\left(1 + r_{\rm P} e^{-i \gamma}\right), \\ \nn
\sqrt{2}\,{\cal A}_{B^-\to \pi^0 K^-}&=& P\,
\Bigg(1+r_{\rm EW}-(r_{\rm T}+r_{\rm C}-r_{\rm P})e^{-i \gamma}+r'_{\rm EW}e^{-i \delta_z}\Bigg), \\ \nn
{\cal A}_{\bar{B}^0\to \pi^+ K^-}&=&P\,
\left(1 -(r_{\rm T}-r_{\rm P}) e^{-i \gamma}\right), \\
\sqrt{2}\,{\cal A}_{\bar{B}^0\to \pi^0 \bar{K}^0}&=&-\,P\,
\Bigg(1-r_{\rm EW}+(r_{\rm C}+r_{\rm P})e^{-i \gamma}-r'_{\rm EW}e^{-i \delta_z}\Bigg).
\eea
Here $r_{\rm T,C,P,EW}$ denote the ratios of the colour-allowed tree-level,
the colour-suppressed tree-level, the doubly Cabbibbo suppressed part of
the QCD penguin and the colour-allowed electroweak penguin amplitudes to
the dominant QCD penguin contribution $P$. The factor $r'_{\rm EW}$
represents the corresponding ratio of the new EW amplitude,
$\delta_z$ being a new weak phase.
Introducing this new term and expanding in the small ratios $r_{\rm T,C,P,EW^{(')}}$,
$\Delta A_{\rm CP}$ and
$S_{\rm CP}(\bar{B}^0\to \pi^0 \bar{K}^0)$ read
\bea\label{5}
\hspace*{-1.0cm}
\Delta A_{\rm CP} &\simeq& -2\left[\IM \left(r_{\rm C}\right)
-\IM \left(r_{\rm T}r_{\rm EW}\right)\right]\sin\gamma
+2 \IM(r'_{\rm EW})\sin\delta_z, \\[0.1cm] \nn
S_{\rm CP}(\bar{B}^0\to \pi^0 \bar{K}^0) &\simeq& \sin 2 \beta
+ 2 \RE \left(r_{\rm C}\right) \cos 2 \beta \sin\gamma
-2 \RE(r'_{\rm EW})  \cos 2 \beta \sin\delta_z.
\eea
The new term can give a large contribution in particular to
$\Delta A_{\rm CP}$ since it is not suppressed by $r_{\rm T}$ as it
happens for the SM EW amplitude $r_{\rm EW}$. On the
other hand, from (\ref{5}) one notes that a similar effect
could be obtained due to an enhanced colour-suppressed tree
contribution $r_{\rm C}$, which in QCD factorization has the largest
uncertainties and is also suggested to be larger by comparison with e.g.
the $\bar{B}\to \pi^0 \pi^0$ decays. Because of these uncertainties,
the problem stays open.

\section{$\bar{B}_s \to \phi \pi$ $\bar{B}_s \to \phi \rho$ and other relevant decays}

In presence of a new EW amplitude, large modifications are expected
in other decays, too. The amplitude of the $\bar{B}_s\to \phi\pi,\phi\rho$
modes reads \cite{QCDF2}
\bea\label{7}\nn
\mathcal A_{\bar B_s \rightarrow \phi M_2} &=&
\frac{A_{\phi M_2}}{\sqrt 2} \bigg( \lambda_u^{(s)} \alpha_2(\phi M_2)
+ \frac{3}{2} \lambda_c^{(s)} \alpha_{\rm 3,EW}(\phi M_2) \\
&&\hspace*{1.3cm}+\,\frac{3}{2}
\lambda_c^{(s)} \left(\delta \alpha_{\rm 3,EW}(\phi M_2)
+\tilde{\alpha}_{\rm 3,EW}(\phi M_2) \right) e^{-i \delta_z}  \bigg),
\eea
with $M_2=\pi,\rho$. We consider explicitly the possibility of having
a new left-handed ($\delta \alpha_{\rm 3,EW}$) and a new right-handed
($\tilde{\alpha}_{\rm 3,EW}$) EW penguin contribution.
In the SM the two contributing terms are the colour-suppressed
tree and the EW penguin amplitude. In QCDF the latter is predicted
to be dominating since the ratio of the two reads e.g.
\be\label{8}
 r_{\phi \pi}^h \equiv \left|\frac{\lambda_u^{(s)}}{\lambda_c^{(s)}}\right|
                     \frac{2}{3}\frac{\alpha_2}{\alpha_{\rm 3,EW}^{c}}
=-0.41^{+0.41}_{-0.37} +  0.13^{+0.30}_{-0.30}\,i.
\ee
The SM branching ratios are quite small \cite{QCDF2},
\be\label{9}
\Br(\bar B_s \rightarrow \pi^0\phi) = 0.15^{+0.11}_{-0.04} \cdot 10^{-6}, \qquad
\Br(\bar B_s \rightarrow \rho^0\phi) = 0.43^{+0.28}_{-0.11} \cdot 10^{-6},
\ee
and a new EW amplitude of the same order as the SM one can easily
enhance the branching fractions up to a factor 4.
\begin{center}
\begin{tabular}{|c|c|c|}
\hline
$\alpha_{\rm3,EW}^{p}+\delta\alpha_{\rm3,EW}^{p}$ & $\tilde{\alpha}_{\rm3,EW}^{p}$ &  \\[0.1cm] \hline
$a_9^{p}+\delta a_9^{p} -a_7^{p}\,-\,\delta a_7^{p}$ &
$-\tilde{a}_9^{p} \,+\,\tilde{a}_7^{p}$ &
if $M_1 M_2=PP$ \\[0.1cm]
$a_9^{p}+\delta a_9^{p} +a_7^{p}+\delta a_7^{p}$ &
 $\phantom{-}\tilde{a}_9^{p} + \tilde{a}_7^{p}$ &
if $M_1 M_2=PV$ \\[0.1cm]
$a_9^{p}+\delta a_9^{p} -a_7^{p}\,-\,\delta a_7^{p}$ &
 $\phantom{-}\tilde{a}_9^{p} \,-\,\tilde{a}_7^{p}$ &
if $M_1 M_2=VP$ \\[0.1cm]
$a_9^{p}+\delta a_9^{p} +a_7^{p}\,+\,\delta a_7^{p}$ &
$-\tilde{a}_9^{p} \,-\,\tilde{a}_7^{p}$ &
if $M_1 M_2=V^0V^0$ \\[0.1cm]
$a_9^{p}+\delta a_9^{p} +a_7^{p}\,+\,\delta a_7^{p}$ &
$-f^{M_1}_\pm\left(\tilde{a}_9^{p} \,+\,\tilde{a}_7^{p}\right)$
& if $M_1 M_2=V^{\pm}V^{\pm}$ \\
\hline
\end{tabular}
\end{center}
A more quantitative prediction can be made only considering some
specific model. This can be understood looking at the table above,
where the EW amplitudes are given in terms of the QCDF building
blocks $a_i$ for decays into $PP$, $PV$, $VP$ amd $VV$ final
states. The left column contains the SM and a possible left-handed
NP contribution, while the right column contains a new right-handed
amplitude. Because of the different interference patterns among the
various terms $a_i$, $\delta a _i$, different NP models can give
very different results, e.g. effects can be larger in $PP$ or $PV$
or $VP$ or $VV$ final states, depending on the handedness of NP,
on its contribution to $\delta a_7$ ($\tilde a_7$) vs. $\delta a_9$
($\tilde a_9$) etc. Because of these patterns, one expects the new
contributions to be relevant in the
$\bar{B}\to \rho\bar{K},\pi\bar{K}^*,\rho\bar{K}^*$ decays, too,
which have the same flavour content as the $\bar{B}\to \pi\bar{K}$
modes. For this reason, we consider them as constraints in our
analysis.

\section{A particular New Physics scenario}

As a first example we considered a modified $Z^0$ scenario, a
well motivated model \cite{Z0mod} which gives rise to new contributions
mainly to the Wilson coefficients of the EW penguin operators. One has 
a $(\bar{s}b)_{V\pm A}$ current mediated at tree level by the $Z^0$ boson.
We write the weak Hamiltonian as
\be\label{10}
{\cal H}^{\rm eff} = \frac{G_F}{\sqrt{2}} \sum_{p=u,c}\lambda_{p}^{(s)}
\bigg(C_1 Q_1^p+  C_2 Q_2^p+\sum_{i=3}^{10}(C_i Q_i+\tilde{C}_i\tilde{Q}_i)\bigg),
\ee
with $\lambda_{p}^{(s)}=V_{pb}V^*_{ps}$. The operators $Q_i$ are defined
as in \cite{QCDF} and the $\tilde{Q}_i$ are obtained from them
replacing $P_L\leftrightarrow P_R$. Parameterizing the flavour
changing $Z^0$-couplings as $\kappa^{sb}_{L,R}\equiv z e^{-i\delta_z}$ as
in the first reference of \cite{Z0mod}, the new contributions to the Wilson
coefficients at the high scale $M_Z$ read
\bea\label{11}\nn
\delta C_3 &=& \frac{\kappa_{L}^{sb}}{6\lambda_t^{(s)}}, \quad\,
\delta C_7 = \phantom{-}\frac{2}{3}\frac{\kappa_L^{sb} \sin^2\theta_W}{\lambda_t^{(s)}}, \quad
\delta C_9 = -\frac{2}{3}\frac{\kappa_L^{sb} \cos^2\theta_W}{\lambda_t^{(s)}},\,\, \\
\tilde C_5 &=& \frac{\kappa_{R}^{sb}}{6\lambda_t^{(s)}}, \quad\,\,\,\,
\tilde C_7 = -\frac{2}{3}\frac{\kappa_R^{sb} \cos^2\theta_W}{\lambda_t^{(s)}}, \quad\,\,
\tilde C_9 = \frac{2}{3}\frac{\kappa_R^{sb} \sin^2\theta_W}{\lambda_t^{(s)}}.
\eea
The model considered is quite general and
simple since we parameterize it with at most two independent free parameters.
Our intention is to consider the experimental results for the $\bar{B}\to \pi \bar{K}$
modes as influenced by this new contributions, so that we fit the free model parameters
to this experimental data. Subsequently we use these results to make predictions
for the $\bar{B}_s\to \phi\pi,\phi\rho$ decays.
However, since the new FCNC coupling $\kappa^{sb}_{L,R}$ contributes also to other
low-energy processes, like e.g. the semileptonic decays $\bar{B}\to X_s l l,X_s \nu\nu$,
these have to be taken into account and they give strong constraints. Once these
constraints are considered, the new coefficients can be at most of the same size as
the SM EW short-distance coefficients, however with a potentially large
weak phase.

\section{Results}

\begin{itemize}
 \item The fit of the current $\bar{B}\to \pi \bar{K}$ data
gives a new EW contribution which is of the same order as the SM
one in case of left-handed NP, while in case of right-handed
NP larger contributions of order 2 to 3 times the SM EW amplitude
are preferred. For comparison, one gets a new contribution of the
same size as the SM EW one for $|\kappa_{L/R}^{sb}|=z=6.9 \cdot 10^{-4}$.
 \item The $VP$, $VV$ modes are more sensitive to right-handed new physics,
which, combined with the previous item, gives in some cases up to
an order of magnitude enhancement for the $\bar{B}_s\to \phi\pi,\phi\rho$ decays.
\item In case of the modified $Z^0$ penguin scenario the
constraints from semileptonic decays limit the possible enhancement
to a factor of 2, which would be difficult to be distinguished from the SM result, 
due to the theoretical error.
\item In case of left-right symmetric NP with
$\kappa_L^{sb}=\kappa_R^{sb}$, the two contributions cancel exactly
in the $\bar{B}_s\to \phi\pi,\phi\rho$ decays and no modifications
arise. However, modifications can still
arise in $\bar{B}\to \pi \bar{K},\pi\bar{K}^*$ decays.
\item The $\bar{B}\to \rho\bar{K},\pi\bar{K}^*,\rho\bar{K}^*,\bar{K}\phi,\bar{K}^*\phi$
modes gives (at present) only weak constraints, mostly in case of right-handed
NP. They are always weaker than the constraints from semileptonic decays.
Their inclusion is however important in other models, where the latter do not apply.
\end{itemize}
  \begin{center}
  \includegraphics[width=0.80\textwidth]{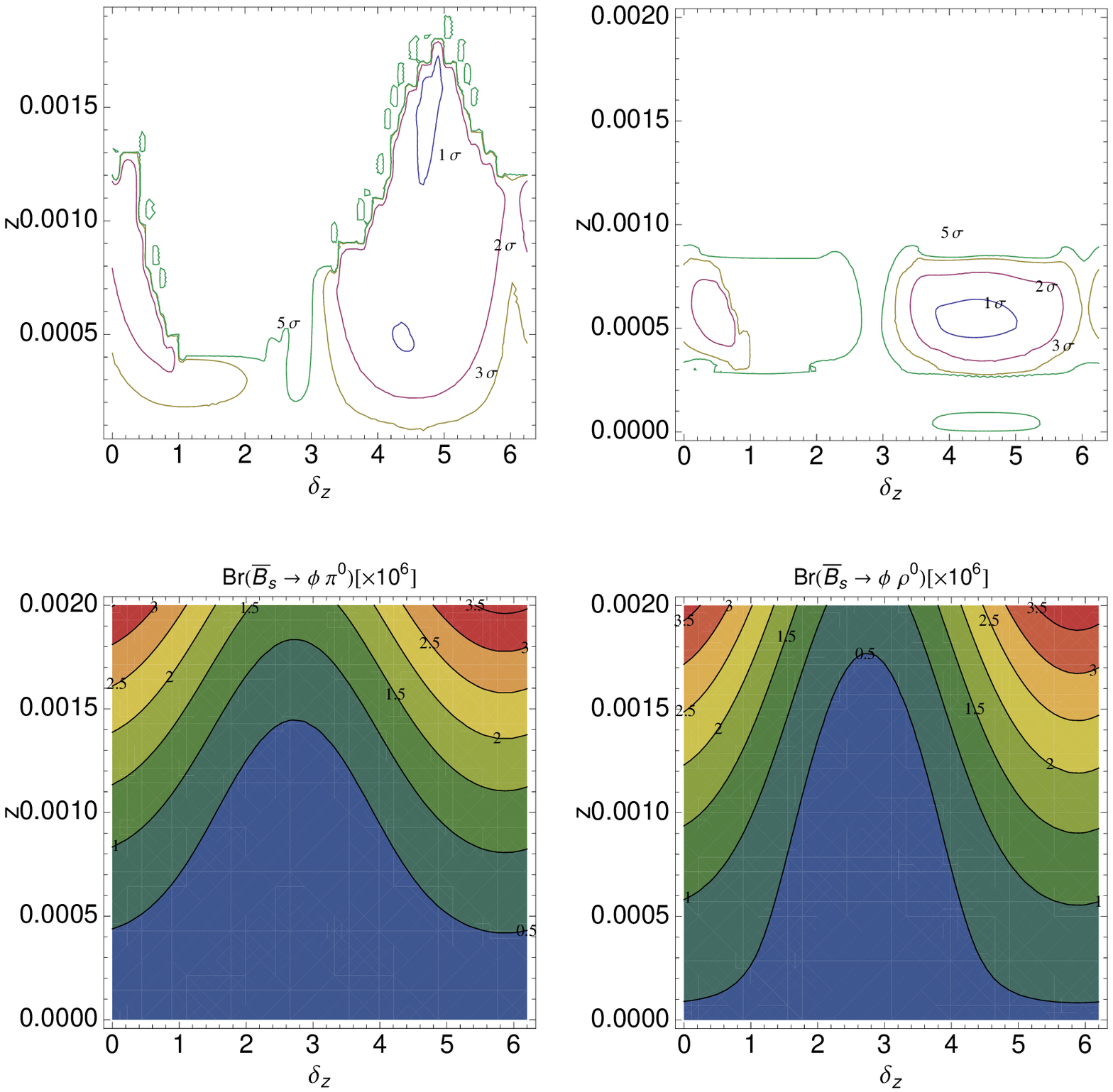}
  \end{center}
In the figure above we support these analyses, providing as an example
the result of the fit (upper graphs) for a right-handed $b\to s$ current,
with (left) and without (right) the constraint from the semileptonic
decay $\bar{B}\to X_s l l$.
The lower graphs show, for the same scenario, the branching ratios
of the $\bar{B}_s\to \phi\pi,\phi\rho$ decays. Inside the $2\,\sigma$
region individuated in the fit, the BRs can be up to one order
of magnitude larger than in the SM.

\section{Conclusion}

Our analysis shows that the decays $\bar{B}_s\to \phi\pi,\phi\rho$
can be used to improve our current understanding of
the $\bar{B}\to \pi \bar{K}$ ``puzzle''. The presence of a new EW
contribution would enhance the branching ratio of the
$\bar{B}_s\to \phi\pi,\phi\rho$ modes. We considered a modified
$Z$ penguin scenario, where the enhancement is limited to a factor
2 by constraints from semileptonic $B$ decays; such an enhancement would 
be difficult to be distinguished from the SM result. However, in other 
extensions of the SM the semileptonic bounds do not apply and one finds enhancement
up to one order of magnitude \cite{Karls}, making them very
interesting decays.
A correlated analysis of these decays, together with the
$\bar{B}\to \rho\bar{K},\pi\bar{K}^*,\rho\bar{K}^*$ modes is useful
to overcome the theoretical uncertainties. We underline therefore that
these decays should be investigated at LHCb and future
super-$B$ factories.

\subsection*{Acknowledgments}

We would like to thank M. Krawczyk, H. Czyz and M. Misiak for organizing
a very stimulating and interesting workshop, U. Nierste for initiating the
project and for useful discussions, and M. Beneke for useful discussions
and suggestions.
This work is supported by the DFG Sonder\-forschungs\-bereich/Transregio~9
``Computergest\"utzte Theoretische Teilchenphysik''

\end{document}